\documentclass[12pt,english]{article}
\usepackage{mathptmx}
\usepackage[T1]{fontenc}
\usepackage[latin9]{inputenc}
\usepackage{geometry}
\usepackage{array}
\usepackage{textcomp}
\usepackage{stmaryrd}
\usepackage{graphicx}

\makeatletter

\providecommand{\tabularnewline}{\\}

\newcommand{\lyxaddress}[1]{
\par {\raggedright #1
\vspace{1.4em}
\noindent\par}
}

\@ifundefined{date}{}{\date{}}
\makeatother

\usepackage{babel}
\begin{document}

\title{An Optimized Bidirectional Quantum Teleportation Scheme with the
use of Bell states}

\author{Mitali Sisodia \thanks{mitalisisodiyadc@gmail.com}}
\maketitle

\lyxaddress{\begin{center}
Department of Physics, Indian Institute of Technology Delhi, \\
New Delhi, Delhi 110016 India. \\
mitalisisodiyadc@gmail.com
\par\end{center}}
\begin{abstract}
Bidirectional quantum teleportation scheme is a two-way quantum communication
process, in which two parties simultaneously receive each other information.
Recently, in paper \cite{2019_six} (Zhou et al., IEEE Access, \textbf{7},
44269 (2019)), a six-qubit cluster state has been used to teleport
three-qubit entangled state (Alice) and arbitrary single-qubit unknown
quantum state (Bob) bidirectionally. In experimental point of view,
the preparation and the maintenance of cluster type entangled state
is very difficult and costly. So, in this paper we have designed an
optimal scheme for bidirectional quantum teleportation scheme and
shown that proposed scheme requires optimized amount of quantum resources. 
\end{abstract}
Controlled NOT (CNOT). Unitary. Measurement. 

\section{Introduction}

In 1993 \cite{bennett}, Bennett et al., proposed a quantum teleportation
scheme for a single-qubit unknown quantum state using two-qubit maximally
entangled state. There are many variant of quantum teleportation (QT)
scheme, in which one variant of QT i.e. bidirectional quantum teleportation
(BQT) attract the attention of the researchers because the beauty
of this scheme is, two parties (Alice and Bob) can get each other
information simultaneously. Later on, various papers on BQT have been
proposed \cite{2019_six,Verma,Hassanpur,sang 2016,chen,asy,swapping,improving,Zadeh,Ail,Kit,Dhara,zhou four,Chaudhary,Kazemik}
(and references therein). In most of the papers, some are mentioned
in Table \ref{tab:References}, they have used highly amount of quantum
resources (multi-qubit entangled states) without giving much attention
to the experimental point of view. Experimental realization and maintenance
of such type of multi-qubit entangled states is very difficulty and
costly. So, these type of schemes demand optimized amount of quantum
resource, which can be prepare easily. In all multi-qubit entangled
state as a quantum resource, Bell state attract much attention because
Bell state is a two-qubit entangled state which is minimum qubit entangled
state and the experimental realization of two-qubit entangled state
is easier and less prone to decoherence as compare with the other
multi-partite entangled states. Keep all these points in mind, I have
designed optimal scheme of BQT for n-qubit quantum state by using
only two Bell states.

The structure of the paper as follows: Section \ref{sec:Optimized-Bidirectional-quantum},
optimized bidirectional scheme for n-qubit quantum state is shown
with an example (shown in subsection \ref{subsec:Proposed-scheme-with}).
Finally, paper is concluded.

\section{Optimized Bidirectional quantum teleportation \label{sec:Optimized-Bidirectional-quantum}}

Alice's $|\psi_{n}\rangle$ and Bob's $|\phi_{n}\rangle$ n-qubit
quantum state of the form of 

\begin{equation}
|\psi_{n}\rangle=A_{1}|\alpha_{1}\alpha_{2}......\alpha_{n}\rangle+A_{2}|\bar{\alpha_{1}}\bar{\alpha_{2}}......\bar{\alpha_{n}}\rangle\label{eq:1}
\end{equation}

\begin{equation}
|\phi_{n}\rangle=B_{1}|\beta_{1}\beta_{2}......\beta_{n}\rangle+B_{2}|\bar{\beta_{1}}\bar{\beta_{2}}......\bar{\beta_{n}}\rangle\label{eq:2}
\end{equation}

Here, $|A_{1}|^{2}+|A_{2}|^{2}=1$ and $|B_{1}|^{2}+|B_{2}|^{2}=1$
. Alice sends $|\psi_{n}\rangle$ to Bob and Bob sends $|\phi_{n}\rangle$
to Alice by using entangled quantum resource. Here, as an optimized
BQT scheme, only two Bell states as a quantum resources is used .
To teleport $|\psi_{n}\rangle$ and $|\phi_{n}\rangle$ to each other
by using optimized amount of quantum resource, Alice and Bob applied
some of CNOT operations. After that teleported state $|\psi_{n}\rangle$
and $|\phi_{n}\rangle$ converted into $|\psi_{n}\rangle^{'}$ and
$|\phi_{n}\rangle^{'}$of the form of 

\[
|\psi_{n}\rangle^{'}=A_{1}|\alpha_{1}\rangle+A_{2}|\bar{\alpha_{1}}\rangle\varotimes|00....\rangle_{n-1}
\]

\begin{equation}
|\psi_{n}\rangle^{'}=|\psi\rangle^{''}\varotimes|00....\rangle_{n-1}\label{eq:3}
\end{equation}

\[
|\phi_{n}\rangle^{'}=B_{1}|\beta_{1}\rangle+B_{2}|\bar{\beta_{1}}\rangle\varotimes|00....\rangle_{n-1}
\]

\begin{equation}
|\phi_{n}\rangle^{'}=|\phi\rangle^{''}\varotimes|00....\rangle_{n-1}\label{eq:4}
\end{equation}

The obtained quantum states $|\psi\rangle^{''}$and $|\phi\rangle^{''}$contains
complete information of $|\psi_{n}\rangle^{'}$ and $|\phi_{n}\rangle^{'}$
and the remaining $\left(n-1\right)$qubits are in $|0\rangle,$ which
are registered qubits. Therefore, the task is reduced to the teleportation
of $|\psi_{n}\rangle^{'}$ and $|\phi_{n}\rangle^{'}$into the teleportation
of $|\psi\rangle^{''}$and $|\phi\rangle^{''}.$ We know that teleportation
of single qubit quantum state of the form of $\alpha|0\rangle+\beta|1\rangle$
required one Bell state which is shown in 1993 by Bennett, so for
$|\psi\rangle^{''}$required one Bell state and other one Bell state
required for $|\phi\rangle^{''}.$ Consequently, two Bell states is
required for our proposed scheme. We can understand easily proposed
n-qubit bidirectional scheme with an example which is shown in below
subsection \ref{subsec:Proposed-scheme-with}.

\begin{table}
\begin{centering}
\begin{tabular}{|c|c|c|c|c|}
\hline 
{\scriptsize{}S.no} & {\scriptsize{}References} & {\scriptsize{}Alice's and Bob's state} & {\scriptsize{}quantum channel} & {\scriptsize{}our proposed scheme}\tabularnewline
\hline 
{\scriptsize{}1.} & {\scriptsize{}(2016) \cite{Hassanpur}} & {\scriptsize{}$a_{0}|00\rangle+a_{1}|11\rangle$} & {\scriptsize{}$\frac{1}{2}\left(|000000\rangle+|000111\rangle+|111000\rangle+|111111\rangle\right)$} & {\scriptsize{}two-Bell states}\tabularnewline
 &  & {\scriptsize{}$b_{0}|00\rangle+b_{1}|11\rangle$} &  & \tabularnewline
\hline 
{\scriptsize{}2. } & {\scriptsize{}(2016) \cite{sang 2016}} & {\scriptsize{}$a_{0}|00\rangle+a_{1}|11\rangle$} & {\scriptsize{}$\frac{1}{2}\left(|00000\rangle+|00111\rangle+|11101\rangle+|111010\rangle\right)$} & {\scriptsize{}two-Bell states}\tabularnewline
 &  & {\scriptsize{}$b_{0}|0\rangle+b_{1}|1\rangle$} &  & \tabularnewline
\cline{1-4} 
{\scriptsize{}5. } & {\scriptsize{} (2017) \cite{asy}} & {\scriptsize{}$a_{0}|000\rangle+a_{1}|111\rangle$} & {\scriptsize{} $\frac{1}{2}\left(|000000000\rangle+|110100001\rangle\right.$} & {\scriptsize{}two-Bell states}\tabularnewline
 &  & {\scriptsize{}$b_{0}|00\rangle+b_{1}|11\rangle$} & {\scriptsize{}$\left.+|001011110\rangle+|111111111\rangle\right)$} & \tabularnewline
\hline 
{\scriptsize{}3.} & {\scriptsize{} (2019) \cite{2019_six}} & {\scriptsize{}$a_{0}|000\rangle+a_{1}|111\rangle$} & {\scriptsize{}$\frac{1}{2}\left(|000000\rangle+|000111\rangle+|111000\rangle+|111111\rangle\right)$} & {\scriptsize{}two-Bell states}\tabularnewline
 &  & {\scriptsize{}$b_{0}|0\rangle+b_{1}|1\rangle$} &  & \tabularnewline
\hline 
{\scriptsize{}4. } & {\scriptsize{}(2020) \cite{chen}} & {\scriptsize{}($a_{0}|0\rangle+a_{1}|1\rangle)$, ($a_{0}^{'}|0\rangle+a_{1}^{'}|1\rangle)$} & {\scriptsize{}$\frac{1}{2\sqrt{2}}\left(|00000000\rangle+|00000011\rangle+|00001100\rangle+|00001111\rangle\right.$} & {\scriptsize{}three-Bell states}\tabularnewline
 &  & {\scriptsize{}$b_{0}|000\rangle+b_{1}|111\rangle$} & {\scriptsize{}$\left.+|00000000\rangle+|00000011\rangle+|00001100\rangle+|00001111\rangle\right)$} & \tabularnewline
\hline 
{\scriptsize{}6.} & {\scriptsize{}(2020) \cite{swapping}} & {\scriptsize{}$a_{0}|00\rangle+a_{1}|11\rangle$} & {\scriptsize{}$\frac{1}{2}\left(|000000\rangle+|000111\rangle+|111000\rangle+|111111\rangle\right)$} & {\scriptsize{}two-Bell states}\tabularnewline
 &  & {\scriptsize{}$b_{0}|00\rangle+b_{1}|11\rangle$} &  & \tabularnewline
\hline 
{\scriptsize{}7.} & {\scriptsize{}(2020) \cite{improving}} & {\scriptsize{}$a_{0}|00\rangle+a_{1}|11\rangle$} & {\scriptsize{}$\frac{1}{2}\left(|00000\rangle+|00111\rangle+|11101\rangle+|11010\rangle\right)$} & {\scriptsize{}two-Bell states}\tabularnewline
 &  & {\scriptsize{}$b_{0}|0\rangle+b_{1}|1\rangle$} &  & \tabularnewline
\hline 
\end{tabular}
\par\end{centering}
\caption{\label{tab:References}In column 4, different type of multi-qubit
quantum state as a resource is used to teleport different type of
unknown quantum states as mentioned in column 3 in previous papers
(column 2). According to our optimized scheme only some of Bell states
is required for the same teleported state (column 3). }
\end{table}

\subsection{Proposed scheme with an example \label{subsec:Proposed-scheme-with}}

In 2019, Zhou et al., proposed a bidirectional scheme \cite{2019_six}
by using six-qubit cluster state of the form of $\frac{1}{2}\left(|000000\rangle+|000111\rangle+|111000\rangle+|111111\rangle\right)_{123456}.$
In this scheme, Alice possesses three-qubit entangled state $|\varphi\rangle_{ABC}=a_{0}|000\rangle+a_{1}|111\rangle,$
which is to be teleported to Bob and Bob possesses an arbitrary single
qubit state $|\varphi\rangle_{D}=b_{0}|0\rangle+b_{1}|1\rangle,$
which is to be teleported to Alice. They have used highly amount of
quantum resource (six-qubit cluster state) and this proposed \cite{2019_six}
bidirectional scheme can be done by using our proposed scheme.\\

\textbf{Improvement of Zhou et al. scheme \cite{2019_six} by using
our proposed scheme: }\\

\begin{figure}
\begin{centering}
\includegraphics[scale=0.38]{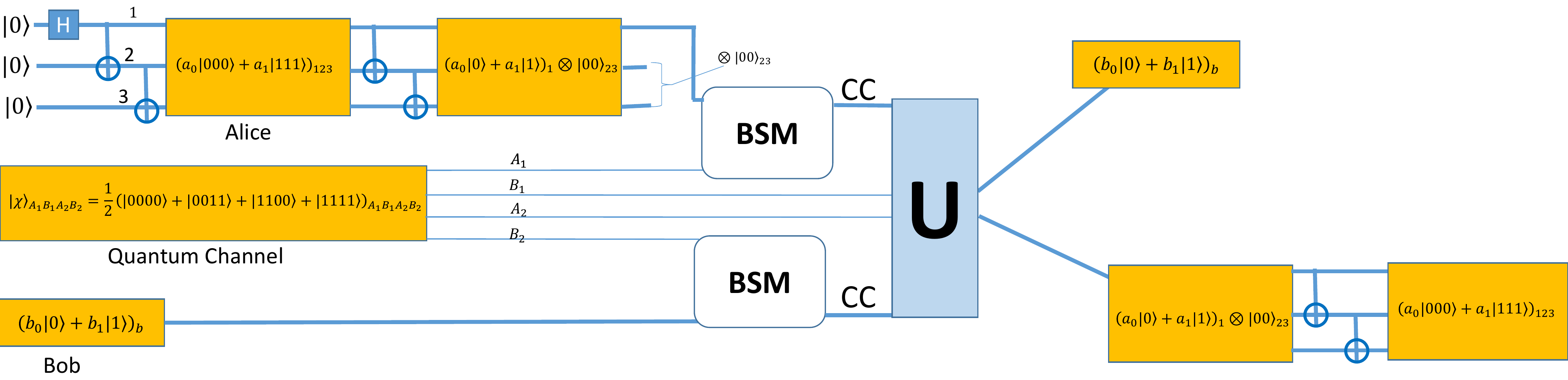}
\par\end{centering}
\caption{\label{fig:Schematic-diagram-of}Schematic diagram of optimized Bidirectional
scheme of Zhou et al., \cite{2019_six} by using Bell states. BSM
corresponds to Bell state measurement, CC represents classical communication
and U represents unitary. }
\end{figure}

Alice wants to teleport three-qubit quantum state $|\varphi\rangle_{ABC}=a_{0}|000\rangle+a_{1}|111\rangle$to
Bob and Bob wish to teleport single qubit state $|\varphi\rangle_{D}=b_{0}|0\rangle+b_{1}|1\rangle$
to Alice. According to our optimized scheme as discussed in Section
\ref{sec:Optimized-Bidirectional-quantum}, Alice applies some CNOT
operations on the last two-qubits in teleported three-qubit entangled
state as shown in Fig. \ref{fig:Schematic-diagram-of}, 

\[
|\varphi\rangle_{ABC}=a_{0}|000\rangle+a_{1}|111\rangle
\]

\begin{equation}
|\varphi\rangle_{a_{1}a_{2}a_{3}}=\left(a_{0}|0\rangle+a_{1}|1\rangle\right)_{a_{1}}\varotimes|00\rangle_{a_{2}a_{3}}=|\Omega\rangle_{a_{1}}\varotimes|00\rangle_{a_{2}a_{3}}\label{eq:5}
\end{equation}

Here, $|\varphi\rangle_{ABC}$ is replace to $|\varphi\rangle_{a_{1}a_{2}a_{3}}$
which corresponds to Alice's qubit and $|\varphi\rangle_{D}$ is replace
with $|\varphi\rangle_{b_{1}}$ which corresponds to Bob's qubit.
In Eq. \ref{eq:5}, $|\Omega\rangle_{a_{1}}$ holds the complete information
of three-qubit entangled state$|\varphi\rangle_{a_{1}a_{2}a_{3}}$,
whereas last two qubits $a_{2}$ and $a_{3}$ are registered qubits.
The task is the telportation of $|\Omega\rangle_{a_{1}}$ state. Now,
Alice sends the unknown single-qubit state to Bob and vice versa.

The quantum channel for this scheme is the tensor product of two Bell
states as

\begin{equation}
\begin{array}{lcl}
|\chi\rangle_{A_{1}B_{1}A_{2}B_{2}} & = & |\psi^{+}\rangle\otimes|\psi^{+}\rangle\\
 & = & \left(\frac{|00\rangle+|11\rangle}{\sqrt{2}}\right)_{AB}\otimes\left(\frac{|00\rangle+|11\rangle}{\sqrt{2}}\right)_{AB}\\
 & = & \left(\frac{|0000\rangle+|0011\rangle+|1100\rangle+|1111\rangle}{2}\right)_{A_{1}B_{1}A_{2}B_{2}}
\end{array}\label{eq:channel}
\end{equation}

Now, the bidirectional scheme for three qubit and single qubit unknown
quantum states is:

\begin{equation}
\begin{array}{lcl}
|\rho\rangle_{a_{1}b_{1}A_{1}B_{1}A_{2}B_{2}} & = & |\Omega\rangle_{a_{1}}\otimes|\varphi\rangle_{b_{1}}\otimes|\chi\rangle_{A_{1}B_{1}A_{2}B_{2}}\end{array}\label{eq:7}
\end{equation}

The complete bidirectional teleportation process of above quantum
state we can find in the Ref. \cite{Verma}, but the difference is
only with the choice of quantum channel, according to our channel,
at the last of the protocol process, we do not need to apply quantum
controlled phase gate operation as used in Ref. \cite{Verma}. After
the bidirectional teleportation of both quantum states $|\Omega\rangle_{a_{1}}$
and $|\varphi\rangle_{b_{1}}$ simultaneously to Bob and Alice, Bob
uses the knowledge of the unitary (CNOT operations) Alice has applied,
he prepares $a_{2}$ and $a_{3}$ qubits in $|0\rangle.$ Then Bob
applies CNOT operations to reconstruct the Alice's three- qubit quantum
state $|\varphi\rangle_{a_{1}a_{2}a_{3}}$. 

To reconstruct unknown quantum state Alice and Bob have to apply some
unitary operations. In above optimized scheme, two copies of Bell
state $|\psi^{+}\rangle=\frac{|00\rangle+|11\rangle}{\sqrt{2}}$ is
used for the teleportation of single qubit states. In given Table
\ref{tab:Combination-of-Bell}, combinations of Bell states as a quantum
channel, measurement operators and correspondingly unitary operations
to reconstruct unknown quantum state is mentioned.
\begin{center}
\begin{table}
\centering{}%
\begin{tabular}{|c|c|c|c|}
\hline 
S.no. & Combination of  & Measurement  & Unitary\tabularnewline
 & Bell states & Operators &  Operations \tabularnewline
\hline 
\hline 
1.  & $|\psi^{+}\rangle\otimes|\psi^{+}\rangle$ & $|\psi^{+}\rangle$, \textbf{$|\psi^{+}\rangle$} & $I,$ $I$\tabularnewline
\hline 
 &  & $|\psi^{+}\rangle,$ $|\psi^{-}\rangle$ & $I,$ $Z$\tabularnewline
\hline 
 &  & $|\psi^{+}\rangle,$ $|\phi^{+}\rangle$  & $I,$ $X$\tabularnewline
\hline 
 &  & $|\psi^{+}\rangle,$ $|\phi^{-}\rangle$ & $I,$ $iY$\tabularnewline
\hline 
 &  & $|\psi^{-}\rangle,$ $|\psi^{+}\rangle$ & $Z,$ $I$\tabularnewline
\hline 
 &  & $|\psi^{-}\rangle$, \textbf{$|\psi^{-}\rangle$} & $Z,$ $Z$\tabularnewline
\hline 
 &  & $|\psi^{-}\rangle,$ $|\phi^{+}\rangle$  & $Z,$ $X$\tabularnewline
\hline 
 &  & $|\psi^{-}\rangle,$ $|\phi^{-}\rangle$  & $Z,$ $iY$\tabularnewline
\hline 
 &  & $|\phi^{+}\rangle,$ $|\psi^{+}\rangle$ & $X,$ $I$\tabularnewline
\hline 
 &  & $|\phi^{+}\rangle$, \textbf{$|\psi^{-}\rangle$} & $X,$ $Z$\tabularnewline
\hline 
 &  & $|\phi^{+}\rangle,$ $|\phi^{+}\rangle$  & $X,$ $X$\tabularnewline
\hline 
 &  & $|\phi^{+}\rangle,$ $|\phi^{-}\rangle$  & $X,$ $iY$\tabularnewline
\hline 
 &  & $|\phi^{-}\rangle,$ $|\psi^{+}\rangle$ & $iY,$ $I$\tabularnewline
\hline 
 &  & $|\phi^{-}\rangle$, \textbf{$|\psi^{-}\rangle$} & $iY,$ $Z$\tabularnewline
\hline 
 &  & $|\phi^{-}\rangle,$ $|\phi^{+}\rangle$  & $iY,$ $X$\tabularnewline
\hline 
 &  & $|\phi^{-}\rangle,$ $|\phi^{-}\rangle$  & $iY,$ $iY$\tabularnewline
\hline 
2. & $|\psi^{-}\rangle\otimes|\psi^{-}\rangle$ & $|\psi^{+}\rangle$, \textbf{$|\psi^{+}\rangle$} & $Z,$ $Z$\tabularnewline
\hline 
 &  & $|\psi^{+}\rangle,$ $|\psi^{-}\rangle$ & $Z,$ $I$\tabularnewline
\hline 
 &  & $|\psi^{+}\rangle,$ $|\phi^{+}\rangle$  & $Z,$ $iY$\tabularnewline
\hline 
 &  & $|\psi^{+}\rangle,$ $|\phi^{-}\rangle$ & $Z,$ $X$\tabularnewline
\hline 
 &  & $|\psi^{-}\rangle,$ $|\psi^{+}\rangle$ & $I,$ $Z$\tabularnewline
\hline 
 &  & $|\psi^{-}\rangle$, \textbf{$|\psi^{-}\rangle$} & $I,$ $I$\tabularnewline
\hline 
 &  & $|\psi^{-}\rangle,$ $|\phi^{+}\rangle$  & $I,$ $iY$\tabularnewline
\hline 
 &  & $|\psi^{-}\rangle,$ $|\phi^{-}\rangle$  & $I,$ $X$\tabularnewline
\hline 
 &  & $|\phi^{+}\rangle,$ $|\psi^{+}\rangle$ & $iY,$ $Z$\tabularnewline
\hline 
 &  & $|\phi^{+}\rangle$, \textbf{$|\psi^{-}\rangle$} & $iY,$ $I$\tabularnewline
\hline 
 &  & $|\phi^{+}\rangle,$ $|\phi^{+}\rangle$  & $iY,$ $iY$\tabularnewline
\hline 
 &  & $|\phi^{+}\rangle,$ $|\phi^{-}\rangle$  & $iY,$ $X$\tabularnewline
\hline 
 &  & $|\phi^{-}\rangle,$ $|\psi^{+}\rangle$ & $X,$ $Z$\tabularnewline
\hline 
 &  & $|\phi^{-}\rangle$, \textbf{$|\psi^{-}\rangle$} & $X,$ $I$\tabularnewline
\hline 
 &  & $|\phi^{-}\rangle,$ $|\phi^{+}\rangle$  & $X,$ $iY$\tabularnewline
\hline 
 &  & $|\phi^{-}\rangle,$ $|\phi^{-}\rangle$  & $X,$ $X$\tabularnewline
\hline 
\end{tabular}
\end{table}
\par\end{center}

\begin{table}
\begin{centering}
\begin{tabular}{|>{\centering}p{0.8cm}|c|>{\centering}p{2.5cm}|>{\centering}p{2.3cm}|}
\hline 
3.  & $|\phi^{+}\rangle\otimes|\phi^{+}\rangle$ & $|\psi^{+}\rangle$, \textbf{$|\psi^{+}\rangle$} & $X,$ $X$\tabularnewline
\hline 
 &  & $|\psi^{+}\rangle,$ $|\psi^{-}\rangle$ & $X,$ $iY$\tabularnewline
\hline 
 &  & $|\psi^{+}\rangle,$ $|\phi^{+}\rangle$  & $X,$ $I$\tabularnewline
\hline 
 &  & $|\psi^{+}\rangle,$ $|\phi^{-}\rangle$ & $X,$ $Z$\tabularnewline
\hline 
 &  & $|\psi^{-}\rangle,$ $|\psi^{+}\rangle$ & $iY,$ $X$\tabularnewline
\hline 
 &  & $|\psi^{-}\rangle$, \textbf{$|\psi^{-}\rangle$} & $iY$ $iY$\tabularnewline
\hline 
 &  & $|\psi^{-}\rangle,$ $|\phi^{+}\rangle$  & $iY,$ $I$\tabularnewline
\hline 
 &  & $|\psi^{-}\rangle,$ $|\phi^{-}\rangle$  & $iY,$ $Z$\tabularnewline
\hline 
 &  & $|\phi^{+}\rangle,$ $|\psi^{+}\rangle$ & $I,$ $X$\tabularnewline
\hline 
 &  & $|\phi^{+}\rangle$, \textbf{$|\psi^{-}\rangle$} & $I,$ $iY$\tabularnewline
\hline 
 &  & $|\phi^{+}\rangle,$ $|\phi^{+}\rangle$  & $I,$ $I$\tabularnewline
\hline 
 &  & $|\phi^{+}\rangle,$ $|\phi^{-}\rangle$  & $I,$ $Z$\tabularnewline
\hline 
 &  & $|\phi^{-}\rangle,$ $|\psi^{+}\rangle$ & $Z,$ $X$\tabularnewline
\hline 
 &  & $|\phi^{-}\rangle$, \textbf{$|\psi^{-}\rangle$} & $Z,$ $iY$\tabularnewline
\hline 
 &  & $|\phi^{-}\rangle,$ $|\phi^{+}\rangle$  & $Z,$ $I$\tabularnewline
\hline 
 &  & $|\phi^{-}\rangle,$ $|\phi^{-}\rangle$  & $Z,$ $Z$\tabularnewline
\hline 
4. & $|\phi^{-}\rangle\otimes|\phi^{-}\rangle$ & $|\psi^{+}\rangle$, \textbf{$|\psi^{+}\rangle$} & $iY,$ $iY$\tabularnewline
\hline 
 &  & $|\psi^{+}\rangle,$ $|\psi^{-}\rangle$ & $iY,$ $X$\tabularnewline
\hline 
 &  & $|\psi^{+}\rangle,$ $|\phi^{+}\rangle$  & $iY,$ $Z$\tabularnewline
\hline 
 &  & $|\psi^{+}\rangle,$ $|\phi^{-}\rangle$ & $iY,$ $I$\tabularnewline
\hline 
 &  & $|\psi^{-}\rangle,$ $|\psi^{+}\rangle$ & $X,$ $iY$\tabularnewline
\hline 
 &  & $|\psi^{-}\rangle$, \textbf{$|\psi^{-}\rangle$} & $X,$ $X$\tabularnewline
\hline 
 &  & $|\psi^{-}\rangle,$ $|\phi^{+}\rangle$  & $X,$ $Z$\tabularnewline
\hline 
 &  & $|\psi^{-}\rangle,$ $|\phi^{-}\rangle$  & $X,$ $I$\tabularnewline
\hline 
 &  & $|\phi^{+}\rangle,$ $|\psi^{+}\rangle$ & $Z,$ $iY$\tabularnewline
\hline 
 &  & $|\phi^{+}\rangle$, \textbf{$|\psi^{-}\rangle$} & $Z,$ $X$\tabularnewline
\hline 
 &  & $|\phi^{+}\rangle,$ $|\phi^{+}\rangle$  & $Z,$ $Z$\tabularnewline
\hline 
 &  & $|\phi^{+}\rangle,$ $|\phi^{-}\rangle$  & $Z,$ $I$\tabularnewline
\hline 
 &  & $|\phi^{-}\rangle,$ $|\psi^{+}\rangle$ & $I,$ $iY$\tabularnewline
\hline 
 &  & $|\phi^{-}\rangle$, \textbf{$|\psi^{-}\rangle$} & $I,$ $X$\tabularnewline
\hline 
 &  & $|\phi^{-}\rangle,$ $|\phi^{+}\rangle$  & $I,$ $Z$\tabularnewline
\hline 
 &  & $|\phi^{-}\rangle,$ $|\phi^{-}\rangle$  & $I,$ $I$\tabularnewline
\hline 
\end{tabular}
\par\end{centering}
\caption{\label{tab:Combination-of-Bell}Combination of Bell states as quantum
channel resource used in bidirectional teleportation scheme with their
measurement operators and corresponding unitary operations applied
by Alice and Bob. Whereas, $|\psi^{+}\rangle=\frac{|00\rangle+|11\rangle}{\sqrt{2}},$
$|\psi^{-}\rangle=\frac{|00\rangle-|11\rangle}{\sqrt{2}},$ $|\phi^{+}\rangle=\frac{|01\rangle+|10\rangle}{\sqrt{2}}$
and $|\phi^{-}\rangle=\frac{|01\rangle-|10\rangle}{\sqrt{2}}$. Unitary
operations $I=\left(\protect\begin{array}{cc}
1 & 0\protect\\
0 & 1
\protect\end{array}\right),$ $X=\left(\protect\begin{array}{cc}
0 & 1\protect\\
1 & 0
\protect\end{array}\right),$ $Y=\left(\protect\begin{array}{cc}
0 & -i\protect\\
i & 0
\protect\end{array}\right)$ and $Z=\left(\protect\begin{array}{cc}
1 & 0\protect\\
0 & -1
\protect\end{array}\right)$}

\end{table}

\section{Conclusion}

In this work, I have optimized the quantum resource for bidirectional
quantum teleportation scheme and shown that only Bell states is required
for the teleportation of any unknown quantum states of the form of
as shown in Eqs. \ref{eq:1} and \ref{eq:2}. It is also shown that
some of previous work on BQT have been used multi-qubit quantum states
as a quantum channel (mentioned in Table \ref{tab:References}), which
is not required according to this proposed scheme. As well as, in
Table \ref{tab:Combination-of-Bell}, also provided the measurement
operators and corresponding their unitary operations for the combination
of Bell states. This work is very useful in the future to teleport
bidirectionally any n-qubit unknown quantum state with the use of
optimal resource. Because the cost and the maintenance of quantum
resource of any scheme is the main concern.


\begin{thebibliography}{10}
\bibitem{2019_six}Zhou, R.G., Xu, R, Lan, H.: Bidirectional quantum
teleportation by using six-qubit cluster state. IEEE Access, 7, 44269
(2019)

\bibitem{bennett}Bennett, C.H., Brassard, G., Cr\textasciiacute epeau,
C., Jozsa, R., Peres, A., Wootters, W.K.: Teleporting an unknown quantum
state via dual classical and Einstein-Podolsky-Rosen channels. Phys.
Rev. Lett. \textbf{70}, 1895 (1993)

\bibitem{Verma}Verma, V., Sisodia, M.: Two-way quantum communication
using four-qubit cluster state: mutual exchange of quantum information.
arXiv preprint arXiv:2107.12169 (2021)

\bibitem{Hassanpur}Hassanpour, S., Houshmand, M.: Bidirectional teleportation
of a pure EPR state by using GHZ states. Quantum Inf. Process., \textbf{15},
905 ( 2016)

\bibitem{sang 2016}Sang, M.H.: Bidirectional quantum teleportation
by using five-qubit cluster state. Int. J. Theor. Phys., \textbf{55},
1333 (2016)

\bibitem{chen}Chen, J., Li, D., Liu, M., Yang, Y.: Bidirectional
quantum teleportation by using a four-qubit GHZ state and two bell
states. IEEE Access, \textbf{8}, 28925 ( 2020)

\bibitem{asy}Choudhury, B.S., Samanta, S.: Asymmetric bidirectional
3\ensuremath{\Leftrightarrow} 2 qubit teleportation protocol between
Alice and Bob via 9-qubit cluster state. Int. J. Theor. Phys., \textbf{56},
3285 (2017)

\bibitem{swapping}Du, Z., Li, X., Liu, X.: Bidirectional quantum
teleportation with ghz states and epr pairs via entanglement swapping.
Int. J. Theor. Phys., \textbf{59}, 622 ( 2020)

\bibitem{improving}Yuan, H., Pan, G.Z.: Improving the Bidirectional
Quantum Teleportation Scheme via Five-qubit Cluster State. Int. J.
Theor. Phys., \textbf{59}, 3387 ( 2020)

\bibitem{Zadeh}Zadeh, M.S.S., Houshmand, M., Aghababa, H.: Bidirectional
teleportation of a two-qubit state by using eight-qubit entangled
state as a quantum channel. Int. J. Theor. Phys., \textbf{56},2101
(2017)

\bibitem{Ail}Aliloute, S., El Allati, A., El Aouadi, I., . Bidirectional
teleportation using coherent states. Quantum Inf. Process., \textbf{20},
1 (2021)

\bibitem{Kit}Kiktenko, E.O., Popov, A.A., Fedorov, A.K., . Bidirectional
imperfect quantum teleportation with a single Bell state. Phys. Rev.
A, \textbf{93}, p.062305 (2016)

\bibitem{Dhara}Choudhury, B.S., Dhara, A.:. A bidirectional teleportation
protocol for arbitrary two-qubit state under the supervision of a
third party. Int. J. Theor. Phys., \textbf{55}, 2275 (2016)

\bibitem{zhou four}Zhou, R.G., Qian, C., Ian, H.: Bidirectional quantum
teleportation of two-qubit state via four-qubit cluster state. Int.
J. Theor. Phys., \textbf{58}, 150 (2019)

\bibitem{Chaudhary}Choudhury, B.S., Samanta, S.: Bidirectional Teleportation
of Three-Qubit Generalized W-States Using Multipartite Quantum Entanglement.
Phys. Part. and Nuclei Lett., \textbf{16}, 206 (2019)

\bibitem{Kazemik}Kazemikhah, P., Aghababa, H.: Bidirectional Quantum
Teleportation of an Arbitrary Number of Qubits by Using Four Qubit
Cluster State. Int. J. Theor. Phys., \textbf{60}, 378 (2021)

\bibitem{Nielsen}Nielsen,M.A., Chuang, I.L.: Quantum Computation
and Quantum Information. Cambridge University Press, Cambridge (2010)

\bibitem{mitali}Sisodia, M., Verma, V., Thapliyal, K., Pathak, A.:
Teleportation of a qubit using entangled non-orthogonal states: a
comparative study. Quantum Inf. Process., \textbf{16}, 76 (2017)
\end{thebibliography}
\end{document}